\documentclass[conference]{IEEEtran}
\IEEEoverridecommandlockouts
\usepackage{cite}
\usepackage{amsmath,amssymb,amsfonts}
\usepackage{graphicx}
\usepackage{textcomp}
\usepackage{xcolor}
\usepackage{graphicx}
\usepackage{subfigure}
\usepackage{amssymb,amsmath}
\usepackage{textcomp}
\usepackage{cite}
\usepackage{float}
\usepackage{bm}
\usepackage{here}
\usepackage{soul, color}
\usepackage{multirow}
\usepackage{booktabs}
\usepackage{multicol}
\usepackage{verbatim}
\usepackage{svg} 
\usepackage{graphicx}
\usepackage{CJKutf8}
\usepackage{amsmath}
\usepackage{algorithm}
\usepackage{algpseudocode}

\usepackage{amsmath}
\usepackage{amssymb,amsmath}
\usepackage{textcomp}
\usepackage{cite}
\usepackage{float}
\usepackage{bm}
\usepackage{here}
\usepackage{soul, color}
\usepackage{multirow}
\usepackage{booktabs}
\usepackage{multicol} 
\usepackage{array}

\usepackage{graphicx}
\usepackage{caption}

\def\BibTeX{{\rm B\kern-.05em{\sc i\kern-.025em b}\kern-.08em
    T\kern-.1667em\lower.7ex\hbox{E}\kern-.125emX}}
\title{Conveniently Identify Coils in Inductive Power Transfer
System Using Machine Learning}

\begin{document}

\author{
    \IEEEauthorblockN{Yifan Zhao$^{1}$, Mowei Lu$^{1}$, Ting Chen$^2$, Heyuan Li$^{3}$, Xiang Gao$^{3}$, Zhenbin Zhang$^2$, Minfan Fu$^3$ and Stefan M. Goetz$^{1*}$}
    \IEEEauthorblockA{\textit{$^1$ Department of Engineering, University of Cambridge, Cambridge, United Kingdom.}}
    \IEEEauthorblockA{\textit{$^2$ School of Information Science and Technology, ShanghaiTech University, Shanghai, China.}}
    \IEEEauthorblockA{\textit{$^3$ School of Electrical Engineering, Shandong University, Jinan, China.}}
    \thanks{This work was supported by National Natural Science Foundation of China under Grant 52477013 and Lingang Laboratory under Grant NO. LG-GG-202402-06-10.}
    \IEEEauthorblockA{\textit{Email: smg84@cam.ac.uk}}
}
\maketitle

\begin{abstract}
High-frequency inductive power transfer (IPT) has garnered significant attention in recent years due to its long transmission distance and high efficiency. The inductance values (\textit{L}) and quality factors (\textit{Q}) of the
transmitting and receiving coils greatly influence the system's operation. Traditional methods involved impedance analyzers or network analyzers for measurement, which required bulky and costly equipment. Moreover, disassembling it for re-measurement is impractical once the product is packaged. Alternatively, simulation software such as HYSS can serve for the identification. Nevertheless, in the case of very high frequencies, the simulation process consumes a significant amount of time due to the skin and proximity effects. More importantly, obtaining parameters through simulation software becomes impractical when the coil design is more complex. This paper firstly employs a machine learning approach for the identification task. We simply input images of the coils and operating frequency into a well-trained model. This method enables rapid identification of the coil's \textit{L} and \textit{Q} values anytime and anywhere, without the need for expensive machinery or coil disassembly.
\end{abstract}

\section{Introduction}
In recent years, IPT has garnered widespread attention due to its ability to overcome the limitations of physical connections. It offers a promising solution for future charging techniques~\cite{huiwang,ipt3,tian2024design,zhao2023detuned,feng2017dual,leigu1,leigu2,minjie1,mowei1,mowei2}. By enabling the wireless transfer of electrical energy, there is a consensus that operating them at very-high frequencies can significantly enhance the transmission distance, reduce system size, and improve overall performance. Increasing the operating frequency of IPT systems has therefore become a widely recognized objective.

The performance of IPT systems is heavily influenced by the \textit{L} and \textit{Q} of the transmitting and receiving coils used, especially in high-frequency cases. These parameters are critical for the efficient operation of the system, i.e., System efficiency and ZVS~\cite{aldhaher2018load,kim2023design,rivas1,rivas2,mingliu1,2,3,4,5,6,7,8,9,10,11}. Also other fields of electronics and power require coils with often intricate design \cite{huang2020quantitative,lucia2013induction, shen2022gradient,pasku2015positioning,9627147,coil1,coil2}. However, measuring these parameters with traditional methods such as impedance or network analyzers can be challenging, as they require expensive and often bulky equipment. Additionally, it is not feasible to disassemble the system for re-measurement once it is manufactured and sealed. Another approach to measure the parameter is to conduct a simulation in software, e.g., HYSS. But in high-frequency cases and when the coil is complex, the simulation will take long time. Therefore, this method becomes impractical.

To solve these issues, this work first proposes a novel approach using machine learning~\cite{ml,li2024unirit,ml3} to identify the \textit{L} and \textit{Q} of coils. By only inputting images of the coils along with their operating frequency into a trained model, we can quickly and accurately determine the necessary parameters without the need for complex and costly measurement devices. This method is portable, easy to use and does not require the coil to be disassembled. Therefore providing a practical solution for measuring coil parameters in various applications. The model uses a convolutional neural network
(CNN) architecture which we trained on a diverse dataset of coils. This dataset includes coils with and without a ferromagnetic core, excitation wires of different thicknesses, and coils of different shapes. Such a comprehensive dataset enables the model to adapt to a wide range of identification tasks. Experimental results have demonstrated that the established model has an identification error rate of only 21.6\%.\footnote{Please note that this work just proposes a novel identification frame using machine learning, the identification error will be reduced by increasing the data set.}
\section{High-Frequency IPT System}

This section describes the basic topology of IPT system and then analyze the influence of \textit{L} and \textit{Q} on operation.

\subsection{Basic Topology of the IPT System}

\begin{figure}[htbp]
    \centering
    \includegraphics[width=0.8\linewidth]{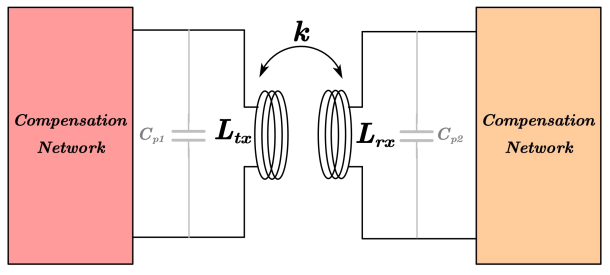}
    \caption{Basic IPT system}
    \label{fig:1}
\end{figure}

Fig.\ \ref{fig:1} illustrates the typical topology of an IPT system. The AC current generated by the inverter is introduced into the primary compensation network. Then it traverses the transmitter coil (Ltx), which transfers energy to the receiver coil (Lrx). Subsequently, the energy is delivered to the load via the secondary compensation network.

\subsection{\textit{L} and \textit{Q} Influence }

\begin{figure}[htbp]
    \centering
    \includegraphics[width=0.95\linewidth]{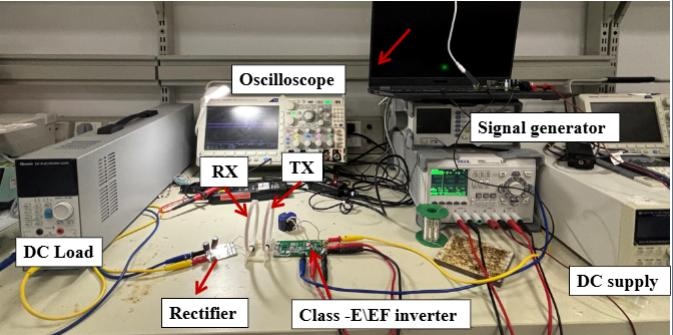}
    \caption{Experimental setup}
    \label{fig:2}
\end{figure}

\begin{figure}[htbp]
    \centering
    \includegraphics[width=0.85\linewidth]{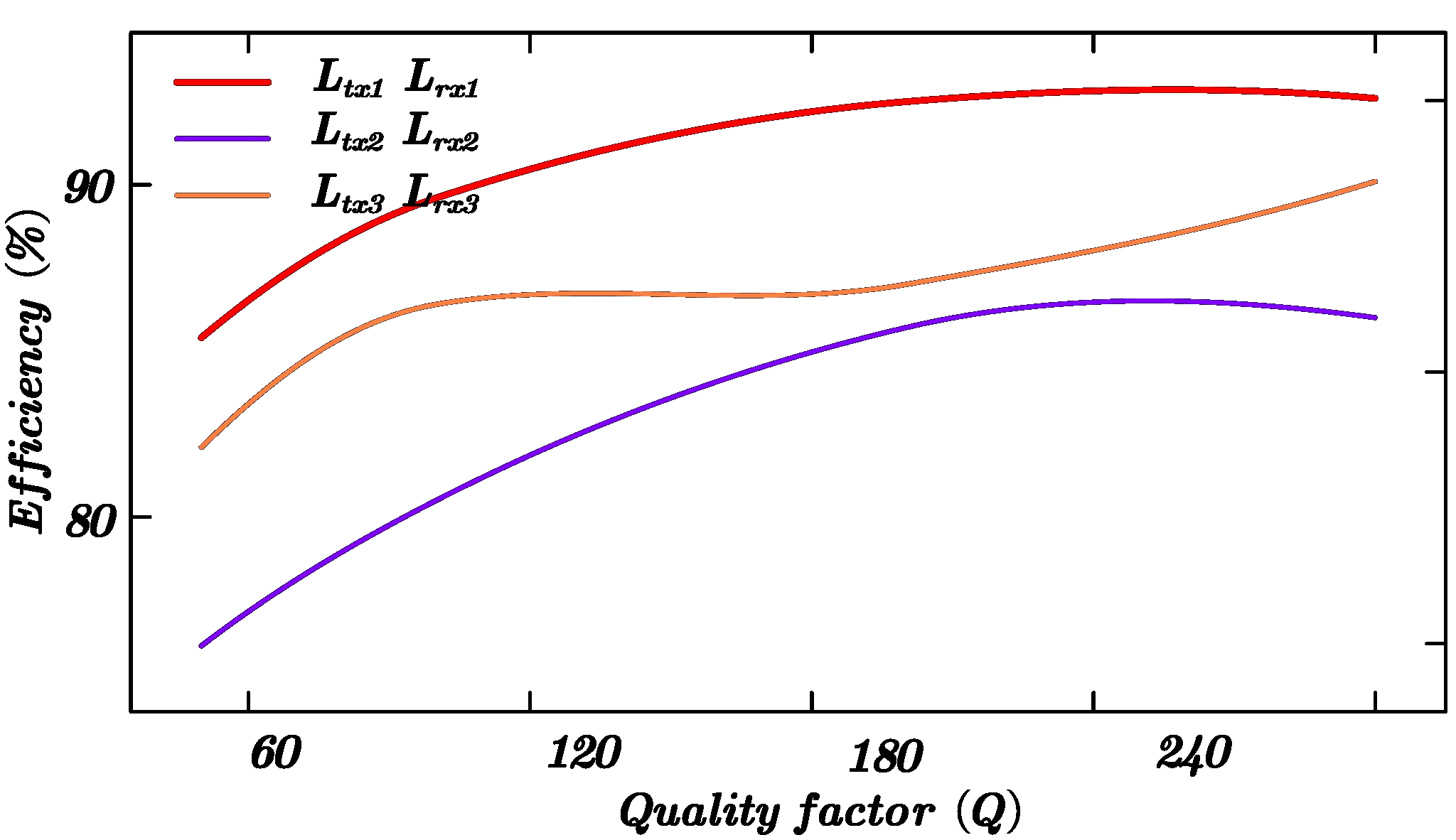}
    \caption{Experimental efficiency}
    \label{fig:3}
\end{figure}

An IPT system prototype operating at 6.78 MHz has been established in Fig.~\ref{fig:2}. The inverter, compensation network, and rectifier respectively use a Class-E inverter, series compensation, and a full bridge. The system's efficiency is measured by altering \textit{L} and quality factor \textit{Q} values of the coils. Fig.~\ref{fig:3} presents the experimental results. It has demonstrated that as \textit{Q} gradually increases, the efficiency also rises. Systems with different \textit{L} values exhibit distinct efficiency curves, thereby confirming the significance of identifying the \textit{L} and \textit{Q} of coils in high-frequency IPT systems.
\section{Identification System Based on CNN Model}
\subsection{CNN Model}
CNNs use convolutional layers that compute only on local pixels and thereby capture local features of the image. Various features can be extracted through multiple convolutional kernels to perform sliding window operations on the input image. These convolutional kernels can detect local information within the image, such as edges, corners, textures, and other distinctive patterns. The convolution formula can be expressed as
\begin{equation}
O(i,j) = \sum_{m=0}^{M-1} \sum_{n=0}^{N-1} I(i+m, j+n) \cdot K(m,n),
\end{equation}
where $O(i,j)$ represents the output feature map at position $(i,j)$, $M$ and $N$ are respectively the height and width of the
convolutional kernel.

A convolutional layer is typically followed by an activation function, with ReLU being the most common one. ReLU aims to introduce nonlinearity, which allows the network to learn more complex patterns. The ReLU function sets
negative input values to zero while it keeps positive values unchanged, which accelerates the training and convergence of the network.

The pooling layer is used to down-sample the feature maps produced by the convolutional layer, reduces computational complexity, and imparts a certain degree of invariance to the features (e.g., translation invariance).
Common pooling methods include max pooling and average pooling. Taking max pooling as an example, the pooling layer selects the maximum value within a small window on the feature map. It forms a down-sampled feature
map, which could be formulated as
\begin{equation}
O(i,j) = Max_{m=0}^{M-1} Max_{n=0}^{N-1} I(i+m, j+n),
\end{equation}
\begin{equation}
O(i,j) = \frac{1}{M \times N} \sum_{m=0}^{M-1} \sum_{n=0}^{N-1} I(i+m, j+n),
\end{equation}
where $M×N$ is the size of the pooling window.

\begin{figure}[htbp]
    \centering
    \includegraphics[width=0.85\linewidth]{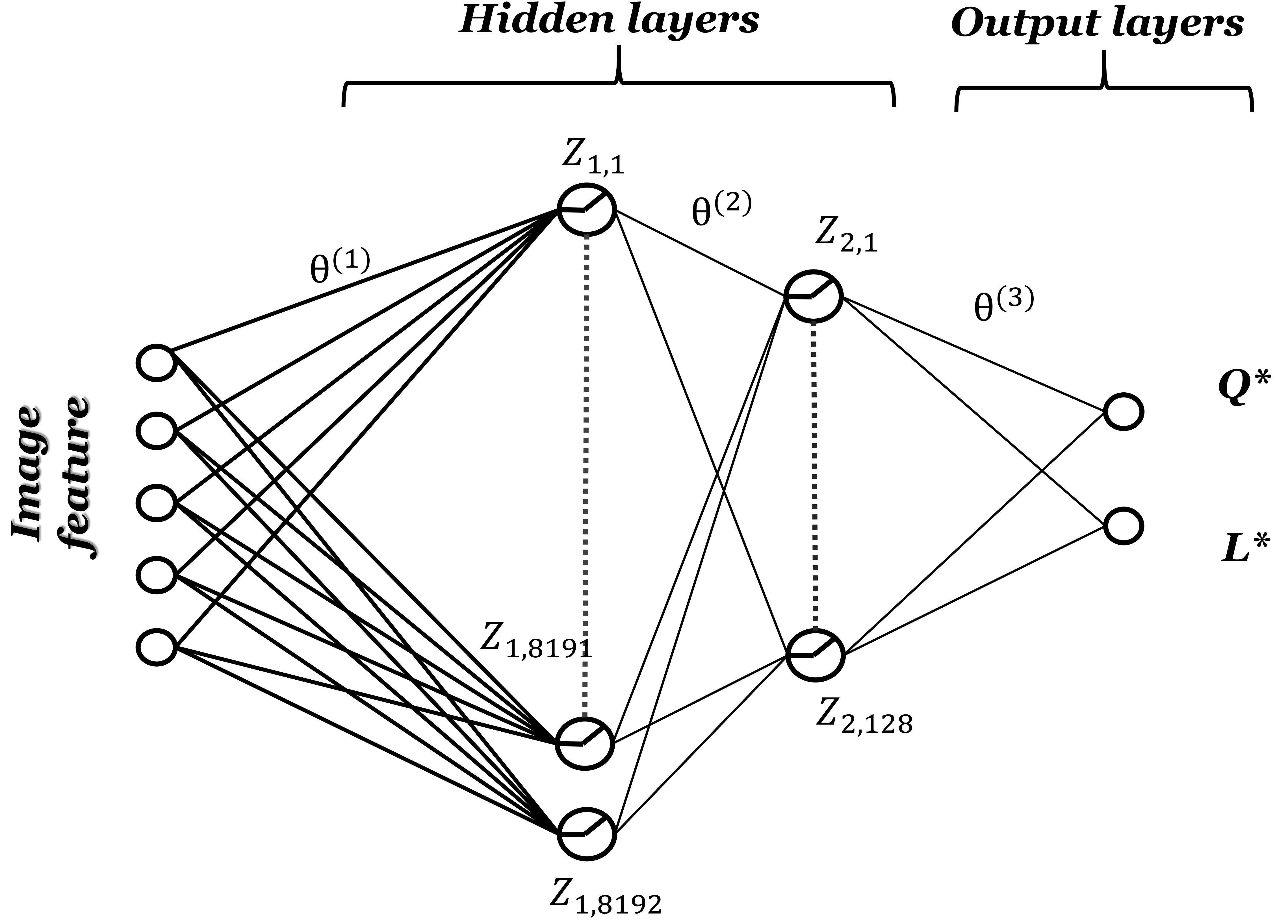}
    \caption{CNN Network}
    \label{fig:4}
\end{figure}

Fig.\ \ref{fig:4} illustrates our decoder module. To ensure the
model's efficiency and effectiveness, we use two fully
connected layers with respective dimensions of 8192 and 128. The features predicted by the CNN are fed into
these fully connected layers, after which the \textit{L} and \textit{Q}
parameters we aim to predict are directly output.

\begin{figure*}[t]
\centering
\includegraphics[width=0.9\textwidth]{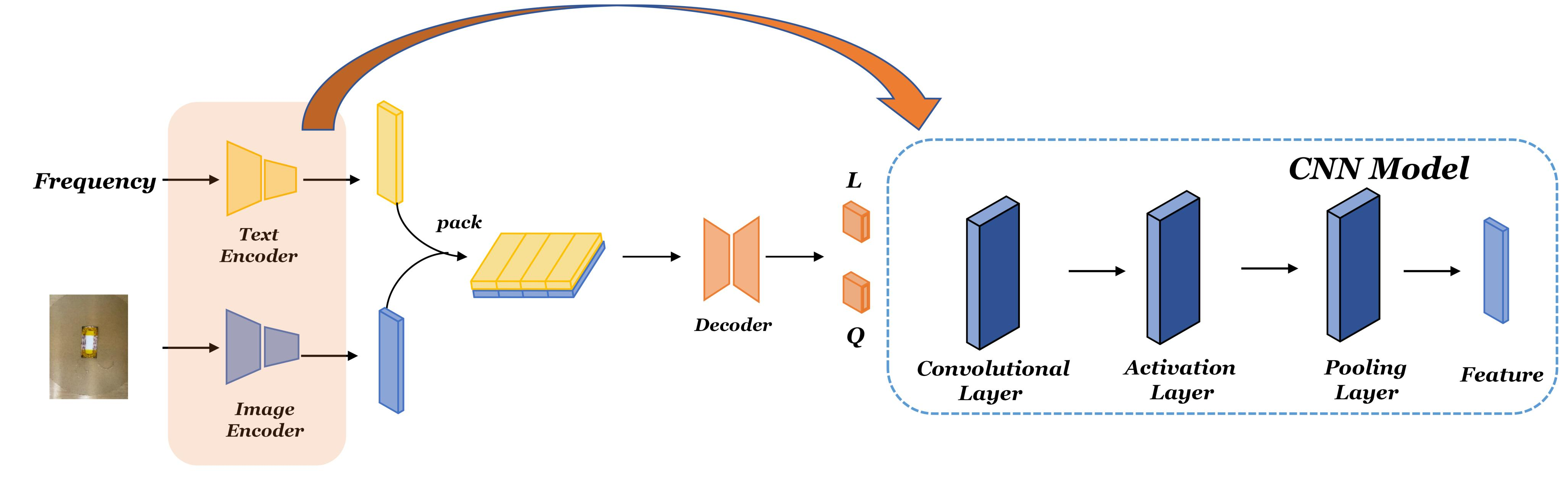}
\caption{Proposed architecture based ML}
\label{fig:5}
\vspace{-10pt}
\end{figure*}
\subsection{Proposed Architecture}
The proposed architecture is shown in Fig.~\ref{fig:5}. By performing convolution operations only on local parts of the image, CNNs are more adept at capturing detailed features such as diameter and coil turns in the coil parameter identification task and improve recognition accuracy. On the other hand, our design incorporates a multi-modal AI representation, as the \textit{L} and \textit{Q} values of the coil are related not only to the coil's parameters but also to the operating frequency of the circuit. Therefore, we simultaneously consider the effect of the operating frequency, as demonstrated in our model diagram. After obtaining the input frequency, we fuse it with the image features to create a multi-modal hybrid feature. Then it is processed by a decoder composed of two fully connected layers to output the resulting \textit{L} and \textit{Q} values. We use the commonly used mean squared error (MSE) as the loss function per
\begin{equation}
MSE = \frac{1}{2} \left[ \left( Q_i - Q_i^* \right)^2 + \left( L_i - L_i^* \right)^2 \right],
\end{equation}
where $Q_i$ represents the identified value of the quality factor, $Q_i^*$  the actual value of the quality factor, $L_i$  the identified value of the inductance value, and $L_i^*$ the actual value of the inductance variable.

 \subsection{Data Collection}
\begin{figure}[htbp]
    \centering
    \includegraphics[width=1\linewidth]{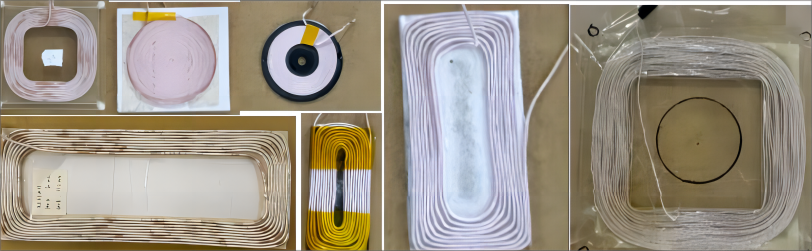}
    \caption{Example images from the dataset}
    \label{fig:6}
\end{figure}
As shown in Fig.~\ref{fig:6}, our dataset is a comprehensive collection that encompasses 20 distinct sets of coils. This diverse assortment includes coils with and without a ferromagnetic core and thereby a broad spectrum of magnetic properties. Furthermore, the dataset incorporates coils with excitation wires of varying thicknesses, which allows the model to discern subtle differences in wire gauge that could impact performance. Additionally, it features coils of diverse shapes, such as circular, rectangular, and irregular forms, which further enhances the model's capability to generalize across different physical configurations.

The meticulously curated and sizable dataset enables the machine learning model to adapt effectively to a wide range of identification tasks, ensuring robustness and versatility in its applications. To gather comprehensive data, we  collected a total of 100 sets of measurements with five different frequencies for each of these 20 sets of coils. This  approach ensures that the model is exposed to a variety of operational conditions and frequencies. It fosters its ability to accurately identify and classify coils under diverse scenarios.

To streamline the processing of the original images and ensure consistency in analysis, we  uniformly resized all twenty images to a standard resolution of 64 $\times$ 64 pixels. This normalization step  maintains uniformity in the input data. 

\subsection{Evaluation metrics}

To evaluate the output quality, we ues two different evaluation metrics, i.e., mean square error
(MSE) and error, which are defined as
\begin{equation}
MSE = \frac{1}{n} \sum_{i=1}^{n} (y_i - \hat{y_i})^2,
\end{equation}
\begin{equation}
error = \frac{1}{2} \left[ \left( Q_i - Q_i^* \right)^2 + \left( L_i - L_i^* \right)^2 \right].
\end{equation}

\section{Experimental Results}
\subsection{Model Training}
\begin{algorithm}
\caption{Coil Parameter Identify}
\begin{algorithmic}[1]
\State \textbf{Data:} $image$, $f$, $Q_{label}$, $\mathcal{I}_{label}$
\State \textbf{Result:} $Q$, $\mathcal{I}$
\State Initialize $Q = 0$, $\mathcal{I}= 0$
\For {$p \in image$ and $N \in f$}
    \State $Q \gets \text{CNN}(image, f_s)$
    \State $\mathcal{I} \gets \text{CNN}(image, f_s)$
    \State \textbf{Calculate MSE Loss:}
        \State $L_Q \gets \text{MSELoss}(Q, Q_{label})$
        \State $L_{\mathcal{I}} \gets \text{MSELoss}(\mathcal{I}, \mathcal{I}_{label})$

    \State \textbf{Update Parameters:}
        \State $\theta_Q \gets \theta_Q - \eta \frac{\partial L_Q}{\partial \theta_Q}$
        \State $\theta_{\mathcal{I}} \gets \theta_{\mathcal{I}} - \eta \frac{\partial L_{\mathcal{I}}}{\partial \theta_{\mathcal{I}}}$
   
\EndFor
\label{AL:1}

\end{algorithmic}
\end{algorithm}

All experiments were implemented with the PyTorch framework on a single GPU (i.e, Nvidia GeForce RTX
4090, 24 GB). The model was trained on M = 16 images  randomly selected from the 20 sets of coil images we obtained. Algorithm 1 details the method. The input to our algorithm consists of an image and a frequency, with the output being \textit{L} and \textit{Q}. First, the input passes through an encoder to extract features, The resulting features are then fed into the decoder, where we employ fully connected layers. The identified values are then compared with the target labels to compute the loss, which is used to update the model parameters and complete one epoch.
\begin{figure}[htbp]
    \centering
    \includegraphics[width=0.95\linewidth]{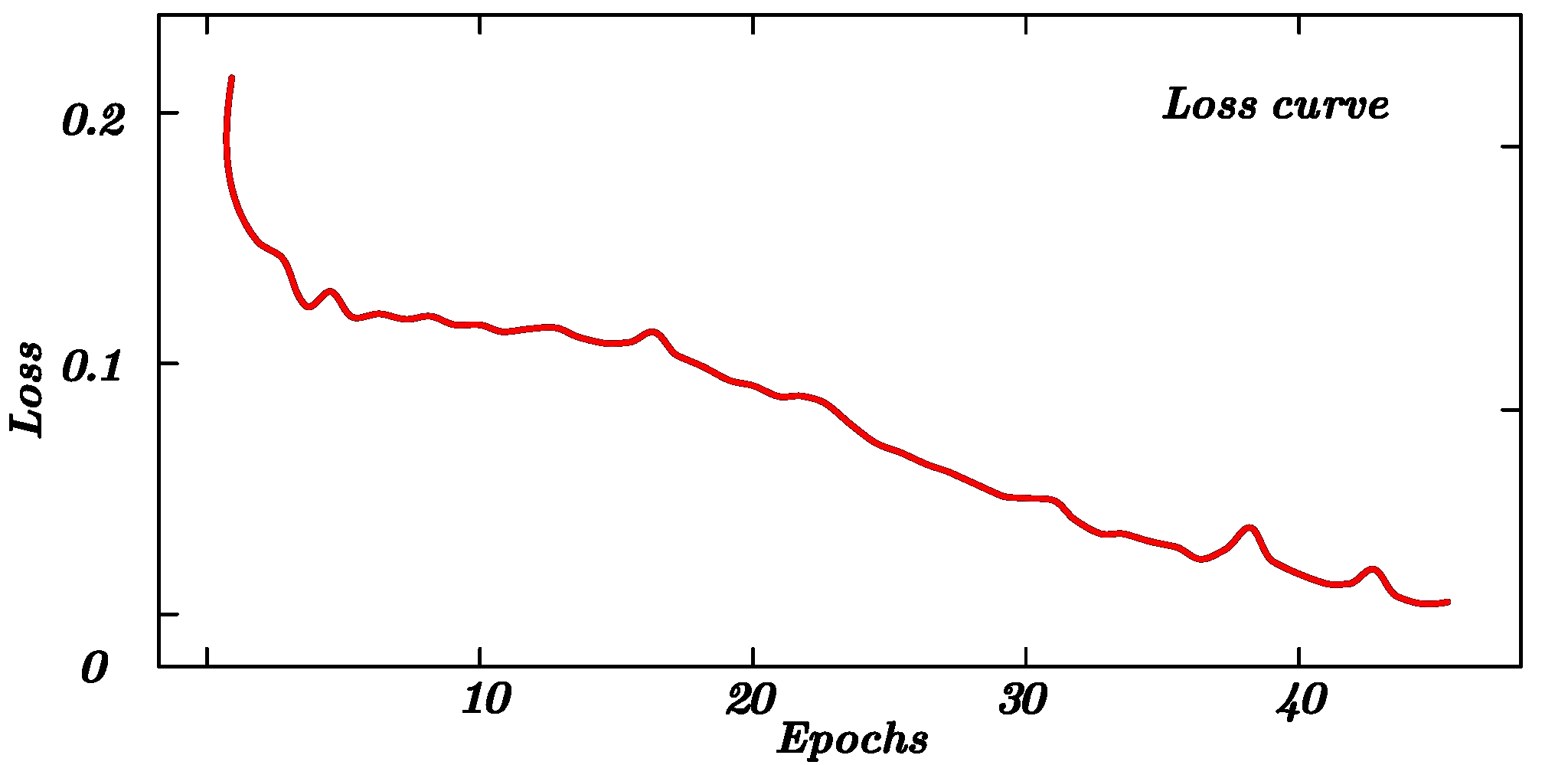}
    \caption{Loss curve}
    \label{fig:7}
\end{figure}
The training loss we obtained is shown in Fig.~\ref{fig:7}. As can be seen, with the increase in the number of epochs, our loss steadily decreases. It has demonstrated the effectiveness of the proposed method.

\subsection{Identification Based on the Proposed Model}
\begin{figure}[htbp]
    \centering
    \includegraphics[width=0.8\linewidth]{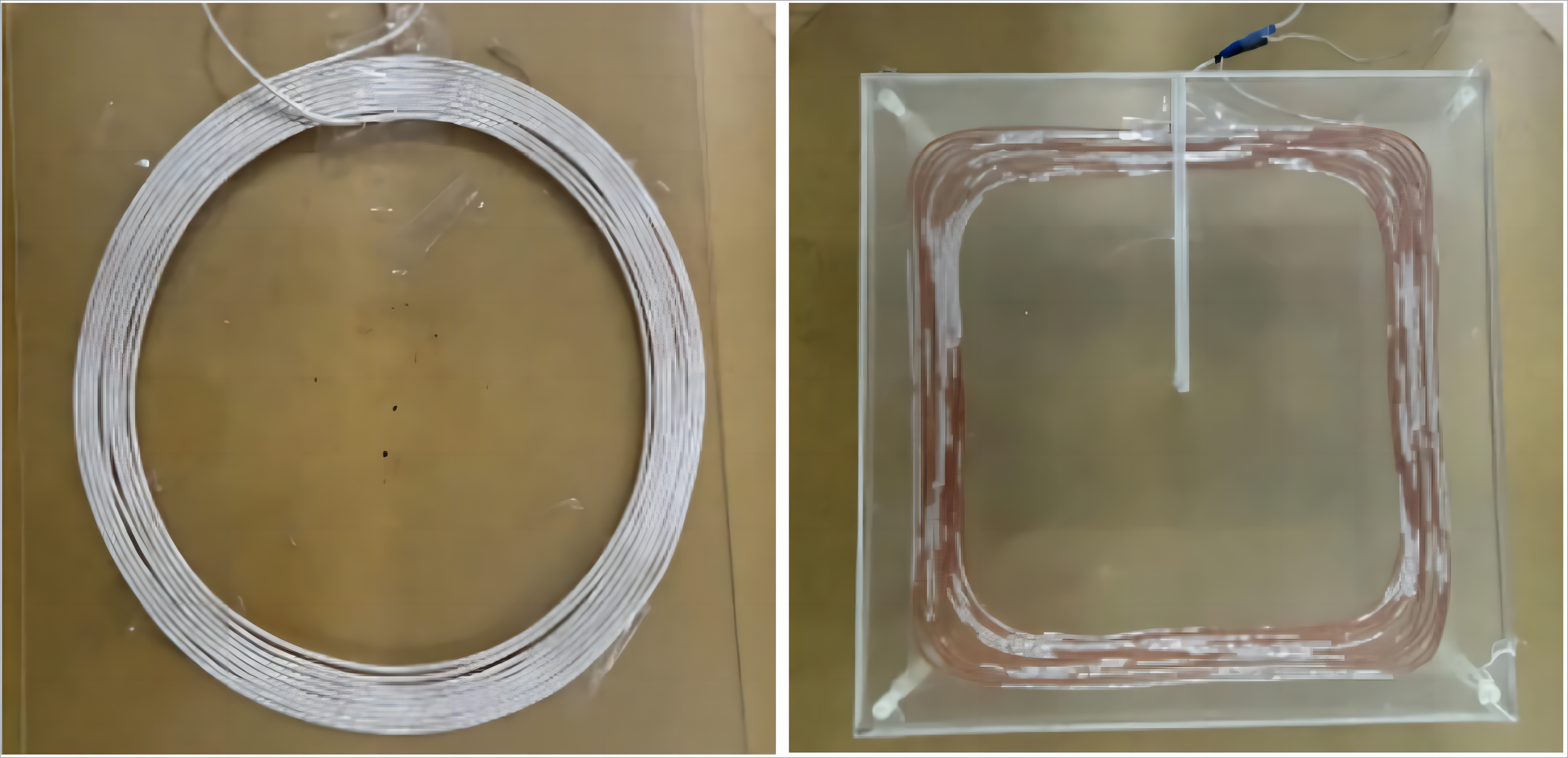}
    \caption{Test coil}
    \label{fig:8}
\end{figure}

\begin{figure}[htbp]
    \centering
    \includegraphics[width=0.98\linewidth]{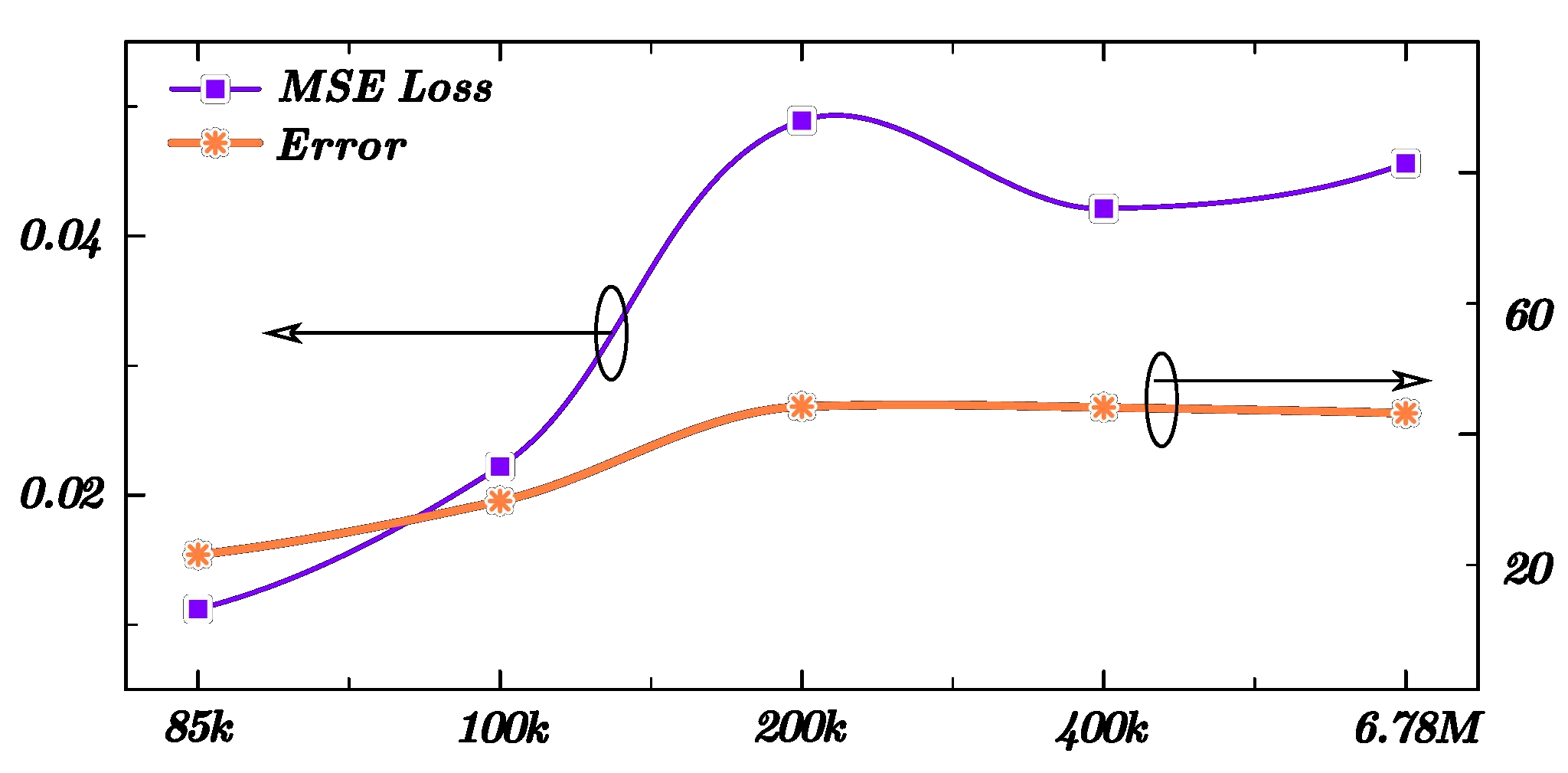}
    \caption{Recognition results}
    \label{fig:9}
\end{figure}

Fig.~\ref{fig:8} shows the experimental test coil, while Fig.~\ref{fig:9} presents the experimental results, which indicate that at 85
kHz, both the MSE loss and error rate reached their lowest values of 0.011\% and 21.6\%, respectively. As the frequency increases, metrics remain at a commendably high standard. The current study only input twenty sets of
coils as the training dataset; if the dataset be expanded, the model's accuracy is expected to significantly improve.
\section{Conclusions}
This paper first introduces a rapid coil parameter recognition technology based on machine learning. Traditional methods based on impedance analyzers and network analyzers are limited by equipment constraints. Moreover, once the entire system is packaged, it is impossible to disassemble it for re-identification. This paper simply recognizes parameters by photographing the coil. Then providing a new paradigm for coil recognition. Experiments have proven that the established model has an identification error rate of only 21.6\%. With the addition of more datasets in the future, the accuracy of the model will be further increased.
\bibliographystyle{IEEEtran}
\bibliography{IEEEabrv,name}

\begin{thebibliography}{10}
\providecommand{\url}[1]{#1}
\csname url@samestyle\endcsname
\providecommand{\newblock}{\relax}
\providecommand{\bibinfo}[2]{#2}
\providecommand{\BIBentrySTDinterwordspacing}{\spaceskip=0pt\relax}
\providecommand{\BIBentryALTinterwordstretchfactor}{4}
\providecommand{\BIBentryALTinterwordspacing}{\spaceskip=\fontdimen2\font plus
\BIBentryALTinterwordstretchfactor\fontdimen3\font minus \fontdimen4\font\relax}
\providecommand{\BIBforeignlanguage}[2]{{%
\expandafter\ifx\csname l@#1\endcsname\relax
\typeout{** WARNING: IEEEtran.bst: No hyphenation pattern has been}%
\typeout{** loaded for the language `#1'. Using the pattern for}%
\typeout{** the default language instead.}%
\else
\language=\csname l@#1\endcsname
\fi
#2}}
\providecommand{\BIBdecl}{\relax}
\BIBdecl

\bibitem{huiwang}
H.~Wang, N.~Tashakor, W.~Jiang, W.~Liu, C.~Q. Jiang, and S.~M. Goetz, ``Hacking encrypted frequency-varying wireless power: Cyber-security of dynamic charging,'' \emph{IEEE Transactions on Energy Conversion}, vol.~39, no.~3, pp. 1947--1957, 2024.

\bibitem{ipt3}
S.~Liu, Y.~Wu, L.~Zhou, R.~Mai, Z.~He, and S.~M. Goetz, ``A two-dimensional misalignment-tolerant ipt system based on three-arm voltage doubler rectifier,'' in \emph{2022 IEEE Energy Conversion Congress and Exposition (ECCE)}, 2022, pp. 1--7.

\bibitem{tian2024design}
X.~Tian, J.~Zhang, H.~Wang, and S.~M. Goetz, ``Design and analysis of automatic modulation and demodulation strategy in wireless power and drive transfer system,'' \emph{IEEE Transactions on Industrial Electronics}, 2024.

\bibitem{zhao2023detuned}
P.~Zhao, J.~Liang, H.~Wang, and M.~Fu, ``Detuned lcc/ss compensation for stable-output inductive power transfer system under ultra-wide coupling variation,'' \emph{IEEE Transactions on Power Electronics}, 2023.

\bibitem{feng2017dual}
H.~Feng, T.~Cai, S.~Duan, X.~Zhang, H.~Hu, and J.~Niu, ``A dual-side-detuned series--series compensated resonant converter for wide charging region in a wireless power transfer system,'' \emph{IEEE Transactions on Industrial Electronics}, vol.~65, no.~3, pp. 2177--2188, 2017.

\bibitem{leigu1}
L.~Gu and J.~Rivas-Davila, ``1.7\,kw 6.78\,mhz wireless power transfer with air-core coils at 95.7\% dc--dc efficiency,'' in \emph{2021 IEEE Wireless Power Transfer Conference (WPTC)}, 2021, pp. 1--4.

\bibitem{leigu2}
L.~Gu, W.~Liang, and J.~R. Davila, ``Design of very-high-frequency synchronous resonant dc-dc converter for variable load operation,'' in \emph{2017 IEEE Energy Conversion Congress and Exposition (ECCE)}, 2017, pp. 3447--3454.

\bibitem{minjie1}
M.~Liu and M.~Chen, ``Dual-band wireless power transfer with reactance steering network and reconfigurable receivers,'' \emph{IEEE Transactions on Power Electronics}, vol.~35, no.~1, pp. 496--507, 2020.

\bibitem{mowei1}
M.~Lu, W.~Mu, M.~Qin, A.~Koehler, J.~Fang, and S.~M. Goetz, ``Differential detection of feeder and mesh impedances through a series–parallel direct-injection soft open point,'' \emph{IEEE Transactions on Power Electronics}, pp. 1--10, 2024.

\bibitem{mowei2}
M.~Lu, M.~Qin, W.~Mu, J.~Fang, and S.~M. Goetz, ``A hybrid gallium-nitride–silicon direct-injection universal power flow and quality control circuit with reduced magnetics,'' \emph{IEEE Transactions on Industrial Electronics}, vol.~71, no.~11, pp. 14\,161--14\,174, 2024.

\bibitem{aldhaher2018load}
S.~Aldhaher, D.~C. Yates, and P.~D. Mitcheson, ``Load-independent class {E/EF} inverters and rectifiers for mhz-switching applications,'' \emph{IEEE Transactions on Power Electronics}, vol.~33, no.~10, pp. 8270--8287, 2018.

\bibitem{kim2023design}
M.~Kim, M.~Jeong, M.~Cardone, and J.~Choi, ``Design of a spiral coil for high-frequency wireless power transfer systems using machine learning,'' \emph{IEEE Journal of Emerging and Selected Topics in Industrial Electronics}, 2023.

\bibitem{rivas1}
J.~Xu, Z.~Tong, and J.~Rivas-Davila, ``1\,kw mhz wideband class {E} power amplifier,'' \emph{IEEE Open Journal of Power Electronics}, vol.~3, pp. 84--92, 2022.

\bibitem{rivas2}
K.~Surakitbovorn and J.~M. Rivas-Davila, ``A simple method to combine the output power from multiple class-e power amplifiers,'' \emph{IEEE Journal of Emerging and Selected Topics in Power Electronics}, vol.~10, no.~2, pp. 2245--2253, 2022.

\bibitem{mingliu1}
S.~Liu, M.~Liu, S.~Yang, C.~Ma, and X.~Zhu, ``A novel design methodology for high-efficiency current-mode and voltage-mode class-e power amplifiers in wireless power transfer systems,'' \emph{IEEE Transactions on Power Electronics}, vol.~32, no.~6, pp. 4514--4523, 2017.

\bibitem{2}
X.~Huang, Y.~Kong, Z.~Ouyang, W.~Chen, and S.~Lin, ``Analysis and comparison of push–pull class-e inverters with magnetic integration for megahertz wireless power transfer,'' \emph{IEEE Transactions on Power Electronics}, vol.~35, no.~1, pp. 565--577, 2020.

\bibitem{3}
H.~Sekiya, K.~Tokano, W.~Zhu, Y.~Komiyama, and K.~Nguyen, ``Design procedure of load-independent class-e wpt systems and its application in robot arm,'' \emph{IEEE Transactions on Industrial Electronics}, vol.~70, no.~10, pp. 10\,014--10\,023, 2023.

\bibitem{4}
C.~H. Lee, G.~Jung, K.~A. Hosani, B.~Song, D.-k. Seo, and D.~Cho, ``Wireless power transfer system for an autonomous electric vehicle,'' in \emph{2020 IEEE Wireless Power Transfer Conference (WPTC)}, 2020, pp. 467--470.

\bibitem{5}
Z.~Yue, Q.~Zhang, Z.~Yang, R.~Bian, D.~Zhao, and B.-Z. Wang, ``Wall-meshed cavity resonator-based wireless power transfer without blocking wireless communications with outside world,'' \emph{IEEE Transactions on Industrial Electronics}, vol.~69, no.~7, pp. 7481--7490, 2022.

\bibitem{6}
M.~Venkatesan, R.~Narayanamoorthi, K.~M. AboRas, and A.~Emara, ``Efficient bidirectional wireless power transfer system control using dual phase shift pwm technique for electric vehicle applications,'' \emph{IEEE Access}, vol.~12, pp. 27\,739--27\,755, 2024.

\bibitem{7}
F.~Lu, H.~Zhang, H.~Hofmann, and C.~C. Mi, ``An inductive and capacitive combined wireless power transfer system with lc-compensated topology,'' \emph{IEEE Transactions on Power Electronics}, vol.~31, no.~12, pp. 8471--8482, 2016.

\bibitem{8}
D.~Ahn and P.~P. Mercier, ``Wireless power transfer with concurrent 200-khz and 6.78-mhz operation in a single-transmitter device,'' \emph{IEEE Transactions on Power Electronics}, vol.~31, no.~7, pp. 5018--5029, 2016.

\bibitem{9}
T.~Mishima and C.-M. Lai, ``Zero-phase-angle load-independent and -adaptable dual-side lcc inductive wireless power transfer system,'' \emph{IEEE Transactions on Transportation Electrification}, vol.~10, no.~2, pp. 3492--3503, 2024.

\bibitem{10}
D.-W. Seo, J.-H. Lee, and H.-S. Lee, ``Optimal coupling to achieve maximum output power in a wpt system,'' \emph{IEEE Transactions on Power Electronics}, vol.~31, no.~6, pp. 3994--3998, 2016.

\bibitem{11}
T.~Fujita, T.~Yasuda, and H.~Akagi, ``A dynamic wireless power transfer system applicable to a stationary system,'' \emph{IEEE Transactions on Industry Applications}, vol.~53, no.~4, pp. 3748--3757, 2017.

\bibitem{huang2020quantitative}
M.-S. Huang, C.-C. Liao, Z.-F. Li, Z.-R. Shih, and H.-W. Hsueh, ``Quantitative design and implementation of an induction cooker for a copper pan,'' \emph{IEEE Access}, vol.~9, pp. 5105--5118, 2020.

\bibitem{lucia2013induction}
O.~Lucia, P.~Maussion, E.~J. Dede, and J.~M. Burd{\'\i}o, ``Induction heating technology and its applications: past developments, current technology, and future challenges,'' \emph{IEEE Transactions on industrial electronics}, vol.~61, no.~5, pp. 2509--2520, 2013.

\bibitem{shen2022gradient}
S.~Shen, N.~Koonjoo, X.~Kong, M.~S. Rosen, and Z.~Xu, ``Gradient coil design and optimization for an ultra-low-field mri system,'' \emph{Applied Magnetic Resonance}, vol.~53, no.~6, pp. 895--914, 2022.

\bibitem{pasku2015positioning}
V.~Pasku, A.~De~Angelis, M.~Dionigi, G.~De~Angelis, A.~Moschitta, and P.~Carbone, ``A positioning system based on low-frequency magnetic fields,'' \emph{IEEE Transactions on Industrial Electronics}, vol.~63, no.~4, pp. 2457--2468, 2015.

\bibitem{9627147}
J.~Zhou, P.~Zhang, J.~Han, L.~Li, and Y.~Huang, ``Metamaterials and metasurfaces for wireless power transfer and energy harvesting,'' \emph{Proceedings of the IEEE}, vol. 110, no.~1, pp. 31--55, 2022.

\bibitem{coil1}
M.~Koehler and S.~M. Goetz, ``A closed formalism for anatomy-independent projection and optimization of magnetic stimulation coils on arbitrarily shaped surfaces,'' \emph{IEEE Transactions on Biomedical Engineering}, vol.~71, no.~6, pp. 1745--1755, 2024.

\bibitem{coil2}
------, ``Simplified {H}1 coil with a single layer of surface conductors,'' \emph{IEEE Access}, vol.~12, pp. 59\,861--59\,867, 2024.

\bibitem{ml}
L.~Jiang and S.~Goetz, ``Artificial intelligence exploring the patent field,'' \emph{arXiv preprint arXiv:2403.04105}, 2024.

\bibitem{li2024unirit}
G.~Li, H.~Cao, M.~Liu, C.~Jiang, and J.~Yang, ``Unirit: Towards few-shot non-rigid point cloud registration,'' \emph{arXiv preprint arXiv:2410.22909}, 2024.

\bibitem{ml3}
A.~Hashemi-Zadeh, N.~Tashakor, M.~Rahnama, C.~T. Touko~Sieyadji, M.~Amirrezai, and S.~Goetz, ``Comparative analysis of model-free deep reinforcement learning controllers for reconfigurable battery systems output voltage regulation,'' in \emph{2024 Energy Conversion Congress \& Expo Europe (ECCE Europe)}, 2024, pp. 1--6.

\end{thebibliography}

\vfill

\end{document}